\documentclass{PoS}
\usepackage{cite}
 \usepackage{amsmath,amssymb}
\usepackage{mathrsfs}
\newcommand{\eref}[1]{eq.~(\ref{#1})}
\newcommand{\cv}{{\cal V}}
\title{Generalized Gradient Flow Equation and Its Applications}

\ShortTitle{Generalized Gradient Flow Equation and Its Applications}
\author{Sinya Aoki\\
        Yukawa Institute for Theoretical Physics, Kyoto University\\
        E-mail: \email{saoki@yukawa.kyoto-u.ac.jp}}
\author{\speaker{Kengo Kikuchi}\\
       Yukawa Institute for Theoretical Physics, Kyoto University\\
       E-mail: \email{kengo@yukawa.kyoto-u.ac.jp}}

\author{Tetsuya Onogi\\
        Osaka University\\
        E-mail: \email{onogi@phys.sci.osaka-u.ac.jp}}
\abstract{We propose a generalization of the gradient flow equation for quantum field theories with non-
linearly realized symmetry. Applying the equation to $\mathcal{N}=1$ $SU(N)$ super Yang-Mills theory in four dimensions, we construct a supersymmetric extension of the gradient flow equation. Choosing an appropriate modification term to damp the gauge degree of freedom, we obtain a gradient flow equation which is closed within the Wess-Zumino gauge. We also apply the equation to the $O(N)$ nonlinear sigma model in two dimensions at large $N$, and show that the two point function in terms of the flowed field is non-perturbatively finite.}

\FullConference{The 33rd International Symposium on Lattice Field Theory\\
		 14 -18 July  2015\\
		 Kobe International Conference Center, Kobe, Japan}

\begin{document}

\section{Introduction}
The gradient flow equation has attracted much attention recently. The equation was originally proposed in the context of the $\mathrm{SU(N)}$ (lattice) gauge theory \cite{Narayanan:2006rf},  \cite{Luscher:2009eq} , \cite{Luscher:2010iy} which was later extended to QCD \cite{Luscher:2013cpa}. The gradient flow has the attractive properties that the expectation value of any gauge invariant local operators in terms of the flowed field are always finite without additional renormalization. In view of this niece property, it is natural question of consider possible extensions of this method for other theories.

However, when we extend the method of the gradient flow to the other systems, we encounter a problem. If the symmetry of the system is nonlinearly realized, the naive gradient flow equation, which means that the derivative of the flowed field is just given of the variation of the action over the field, does not respect the symmetry.

In this proceedings, we report our proposal for the generalization of the gradient flow equation with nonlinearly realized symmetry and our study on its application to some examples of field theories. Using this equation, we can construct the various type of gradient flow equation for quantum field theories whose symmetry is non-linearly realized.  
We then apply the generalized gradient flow equation to two systems. One is the $\mathcal{N}=1, d=4$, $SU(N)$ super Yang-Mills theory \cite{Kikuchi:2014rla}, and the other is the $d=2, O(N)$ nonlinear sigma model \cite{Aoki:2014dxa}. In the case of the super Yang-Mills theory, we construct a natural extension of the gradient flow using the superfield formalism. With a special choice of the modification term in the gradient flow equation, we obtain a closed equation within the Wess-Zumino(WZ) gauge.
 We also find an exact solution of the generalized gradient flow equation for the $O(N)$ nonlinear sigma model in the large $N$ limit, which shows the finiteness of the two point function in terms of the flowed field non-perturbatively. 

\section{Generalized Gradient Flow Equation}

The generalized gradient flow equation  proposed in Ref.\cite{Kikuchi:2014rla} 
keeps the nonlinearly realized symmetry of the system, in other words,  the time evolution and the symmetry transformation  commute.
Let us first define the norm of the variation of fields $\delta \varphi(x)$ as
\begin{eqnarray}
||\delta \varphi||^2=\int d^D x g_{ab} (\varphi(x))\delta \varphi^a (x) \delta \varphi^b (x),~~~~~a=1, 2,\cdots, M,
\label{eq:norm_phi}
\end{eqnarray}
where $M$ is the number of components of the field and $g_{ab}(\varphi(x))$ is the metric in the functional space.
The metric should be chosen in such a way that the norm is invariant under the symmetry transformation. 
The generalized gradient flow equation is then provided as
\begin{eqnarray}
\frac{\partial \phi^a_t(x)}{\partial t}=-g^{ab}(\phi_t(x))\frac{\delta S(\varphi)}{\delta \varphi^b}|_{\varphi\rightarrow\phi}\label{ggfe}
\end{eqnarray}
where the $g^{ab}$ is inverse matrix of the metric $g_{ab}$. 
We note that for linearly realized symmetry, the metric $g_{ab}$ is trivial  so that the generalized gradient flow equation reduces to the naive gradient flow equation.

Whether one can find an appropriate metric or not for a given field theory is quite nontrivial, 
 but there are a few examples where one can find the metric explicitly as we explain in Sec.3 and Sec.4.

\section{$N=1$, $d=4,$ $SU(N)$ Super Yang-Mills Theory}

\subsection{Gradient Flow Equation of Super Yang-Mills Theory}
The original gradient flow equation for $SU(N)$Yang-Mills theory \cite{Luscher:2010iy} is given by 
\begin{eqnarray}
\frac{\partial B^a (x)}{\partial t}&=&-\delta^{ab}\frac{\delta S_{\mathrm{YM}}}{\delta B^b(x)}+\alpha_0 \delta B^a(x),\label{YM}\\
B_\mu |_{t=0}&=&A_\mu.
\end{eqnarray}
The first term of the RHS of \eref{YM} is the gradient of the Yang-Mills action. The second term is the modification term introduced to suppress the gauge degrees of freedom. This term has to be proportional to gauge transformation so that it does not affect physical quantities.

We extend this equation to supersymmetric one by the following replacement  of the Yang-Mills gradient flow equation.
\begin{itemize}
\item Yang-Mills action $S_{\mathrm{YM}}$ $\rightarrow$ Super Yang-Mills action $S_{\mathrm{SYM}}$,
where
\begin{eqnarray}
S_{\mathrm{SYM}}&=&-\int d^4 x \int d^2\theta \mathrm{Tr}[W^\alpha W_\alpha]+h.c.,\\
W_{\alpha}&=&-\bar{D}\bar{D}(e^{-V}D_{\alpha}e^V).
\end{eqnarray}
\item Gauge field $A_\mu(x)$  $\rightarrow$ Superfield $V(z)$, where the argument $z$ stands for super coordinate $(x, \theta, \bar{\theta})$. 
\item New gauge field $B_\mu(t, x)$ $\rightarrow$ New superfield $\cv (t, z)$. The component of superfield $\cv$ is defined by ${\cal V}=\{c,\chi,\bar{\chi},m,m^*,v_m,\lambda, \bar{\lambda},d\}$. We impose the initial condition $\cv(0, z)=V(z)$.
\item Gauge transformation $\delta B^a(x)$ $\rightarrow$ Super gauge transformation $\delta \cv^a (x)$ defined by 
$\delta \cv=L_{\cv/2}\cdot [(\Phi-\Phi^{\dagger})+\mathrm{coth}(L_{\cv/2})\cdot(\Phi+\Phi^{\dagger})]$, where $\Phi$ is a chiral superfield.
\item Metric $\delta^{ab}$ $\rightarrow$ $g^{ab}(\cv)$   
\end{itemize}
Then we obtain the general form of the supersymmetric extension of the gradient flow equation,
\begin{eqnarray}
\frac{\partial \cv^a}{\partial t}=-g^{ab}(\cv)\frac{\delta S_{\mathrm{SYM}}}{\delta \cv^b}+\alpha_0\delta \cv^a\label{general1},
\end{eqnarray}
 where 
 \begin{eqnarray}
g^{ab}(\cv)&=&-4\mathrm{Tr}\left[\left(\frac{L_{\cv}}{1-e^{-L_{\cv}}}\cdot T^a\right)\left(\frac{L_\cv}{1-e^{-L_\cv}}\cdot T^b\right)\right], \label{met2}\\
L_{\cv}\cdot&=&[\cv, \cdot].
\end{eqnarray}
 Substituting the explicit forms of $g^{ab}(\cv), \frac{\delta S_{\mathrm{SYM}}}{\delta \cv^b}$ into \eref{general1}, $\delta \cv_a$, we obtain the gradient flow equation in the matrix form as 
\begin{eqnarray}
\frac{\partial \cv}{\partial t}
&=&\frac{L_\cv}{1-e^{-L_\cv}}(F+\alpha_0\Phi_\cv)+h.c.,\label{81}
\end{eqnarray}
where
\begin{eqnarray}
F=D^\alpha w_\alpha+\{e^{-\cv}D^\alpha e^{\cv}, w_\alpha\}.
\end{eqnarray}
and $\Phi_\cv$ is a chiral field, $\cv=\cv^a T^a$ and $T^a$ is a representation matrix. The field strength $w_\alpha$ is given by $w_\alpha \equiv -\bar{D}\bar{D}(e^{-\cv}D_\alpha e^{\cv})$.

\subsection{Wess-Zumino Gauge}
The gradient flow equation \eref{81} has infinite number of commutators in the general gauge, which makes it difficult to solve. In order to obtain the flow equation with finite number of terms, we chose the WZ gauge. However, generally the time evolution from the flow equation can carry the system away from the WZ gauge. Therefore, the most important question is whether there exists the special chiral field $\Phi_\cv$ which give the super gauge transformation keeping the WZ gauge. As a result, we find that such a $\Phi_\cv$ exists as follows, 
\begin{eqnarray}
\alpha_0&=&1,\\
\delta \cv&=&\Phi_\cv+\Phi_\cv^{\dagger}+\frac{1}{2}[\cv, \Phi_\cv-\Phi_\cv^{\dagger}]+\frac{1}{12}[\cv, [\cv,\Phi_\cv+\Phi_\cv^{\dagger}]],
\end{eqnarray}
where\begin{eqnarray}
\Phi_\cv&=&\bar{D}^2(D^2\cv+[D^2 \cv, \cv]).\label{important}
\end{eqnarray}

Finally, we obtain the flow equations for the each component of the vector multiplet as
{\allowdisplaybreaks
\begin{eqnarray}
\dot{c}&=&0, \dot{\chi}=0, \dot{\bar{\chi}}=0, \dot{m}=0, \dot{m}^*=0,\\
\dot{v}_m&=&-16\mathscr{D}^k v_{mk}+16\mathscr{D}_m\partial_k v^k-8\{\bar{\lambda}_{\dot{\alpha}},(\bar{\sigma}_m \lambda)^{\dot{\alpha}} \},\\
\dot{\bar{\lambda}}&=&-16\bar{\sigma}^k\sigma^m\mathscr{D}_k\mathscr{D}_m\bar{\lambda}+8[\bar{\lambda}, d+i\partial_m v^m],\label{matter1}\\
\dot{\lambda}&=&-16\sigma^k\bar{\sigma}^m\mathscr{D}_k\mathscr{D}_m\lambda-8[\lambda, d-i\partial_m v^m],\label{matter2}\\
\dot{d}&=&16\Box d+16i[v_m, \partial^m d]\nonumber\\
&&+2i\mathrm{Tr}[\bar{\sigma}^m\sigma^l\bar{\sigma}^n\sigma^k-\bar{\sigma}^m\sigma^k\bar{\sigma}^n\sigma^l]\mathscr{D}_n\mathscr{D}_l v_{mk}\nonumber\\
&&+8i\{\bar{\lambda}_{\dot{\alpha}}, (\bar{\sigma}^m\mathscr{D}_m\lambda)^{\dot{\alpha}}\}-8i\{\lambda^{\alpha}, (\sigma^m\mathscr{D}_m\bar{\lambda})_{\alpha}\}\nonumber\\
&&-4[v_m, [v^m, d]].   
\end{eqnarray}}
We find that the flow equations for each component are consistent with WZ gauge. Here we choose initial conditions to satisfy the WZ gauge at $t=0$ as
{\allowdisplaybreaks
\begin{eqnarray}
{c}|_{t=0}&=&0, {\chi}|_{t=0}=0, {\bar{\chi}}|_{t=0}=0, {m}|_{t=0}=0, {m}^*|_{t=0}=0,\nonumber\\
{v}_m|_{t=0}&=&V_m, {\bar{\lambda}}|_{t=0}=\bar{\Lambda}, {\lambda}|_{t=0}=\Lambda, {d}|_{t=0}=D.
\end{eqnarray}}

\section{d=2 O(N) Nonlinear Sigma Model}

\subsection{Gradient Flow Equation of d=2 O(N) Nonlinear Sigma Model}

We consider the $O(N)$ nonlinear sigma model in two dimensions. The action is given by
\begin{eqnarray}
S=\frac{1}{2g^2} \int d^2x \sum_{a,b=1}^{N-1}g_{ab}(\varphi)\left(\partial_\mu \varphi^a \partial_\mu\varphi^b \right)
\end{eqnarray}
where the metric $g_{ab}$ and its inverse $g^{ab}$ are provided by
\begin{eqnarray}
g_{ab}(\varphi) = \delta_{ab} + \frac{\varphi^a \phi^b}{1-\displaystyle{ (\varphi^c)^2}}, &
g^{ab}(\varphi)=\delta_{ab}-\varphi^a\phi^b.
\end{eqnarray}
Introducing the momentum cutoff $\Lambda$, the gap equation leads to
\begin{eqnarray}
1=\frac{\lambda}{4\pi}\ln\frac{\Lambda^2+m^2}{m^2},
\end{eqnarray}
where $\lambda\equiv g^2 N$ is the 't Hooft coupling constant, and $m$ is the dynamically generated mass.

Using \eref{ggfe}, we obtain the gradient flow equation of the $O(N)$ nonlinear sigma model in two dimensions as
\begin{eqnarray}
\frac{d}{dt}\phi^a=\square\phi^a+\phi^a\partial_\mu\vec{\phi}\cdot\partial_\mu\vec{\phi}+\frac{\phi^a (\partial_\mu \vec{\phi}^2)^2}{4(1-\vec{\phi}^2)},\label{eq:Gflow}
\end{eqnarray}
where the initial condition is $\phi^a(0, x)=\varphi(x)$. Here we rescaled $t$ as $t\rightarrow g^2 t$ and summations over indices are implicitly assumed as
\begin{eqnarray}
 \vec{\phi}^2 = \sum_{b=1}^{N-1} (\phi^b)^2, & &
(\partial_\mu \vec{\phi})^2 = \displaystyle{\sum_{b=1}^{N-1}}(\partial_\mu \phi^b)^2 .
\end{eqnarray}

\subsection{Solution to Gradient Flow Equation in Large N Expansion}
To solve \eref{eq:Gflow},
we employ the large $N$ expansion, which makes the equation much simpler
after dropping the subleading contributions. We take the ansatz for the solution to the gradient flow equation as 
\begin{eqnarray}
\phi^a(t,p) &=& f(t) e^{-p^2 t} \sum_{n=0}^\infty : X_{2n+1}^a(\varphi, p, t) :  \ ,
\label{eq:sol}
\end{eqnarray}
where $X_{2n+1}$ only contains $2n+1$-th order of $\varphi$, 
and $: {\cal O}: $ represents the "normal ordering", which prohibits 
self-contractions within  the operator ${\cal O}$. Formally we can define the normal ordering recursively in the perturbation theory around the large $N$ vacuum as
\begin{eqnarray}
:\varphi^a(p): &=& \varphi^a(p) \\
\langle :\varphi^{a_1}(p_1) \varphi^{a_2}(p_2): {\cal O} \rangle &=& \langle \varphi^{a_1}(p_1) \varphi^{a_2}(p_2) {\cal O}\rangle - \langle \varphi^{a_1}(p_1) \varphi^{a_2}(p_2)\rangle \langle {\cal O}\rangle\\
\langle : \prod_{j=1}^n \varphi^{a_j}(p_j) : {\cal O} \rangle &=& \langle  \prod_{j=1}^n \varphi^{a_j}(p_j)  {\cal O} \rangle
-\sum_{k\not= l}^n \langle \varphi^{a_k}(p_k) \varphi^{a_l}(p_l)\rangle  \langle  : \prod_{j\not=k,l}^{n-2} \varphi^{a_j}(p_j):  {\cal O} \rangle
\end{eqnarray}
for an arbitrary operator ${\cal O}$.
From the initial condition for $\varphi$, we have 
\begin{eqnarray}
X_1^a(\varphi,p, 0) &=& \varphi^a(p), \qquad f(0)=1, \qquad X_{2n+1}^a(\varphi,p,0) = 0, \ n \ge 1 .
\end{eqnarray}

The gradient  flow equation in the momentum space is written as
\begin{eqnarray}
L^a(t,p) &\equiv & \dot\phi^a(t,p) + p^2 \phi^a(t,p) \nonumber \\
=R^a(t,p)&\equiv& - \int_{p}^{3}  \phi^a(t,p_1) (p_2\cdot p_3) \vec\phi(t,p_2)\cdot\vec\phi(t,p_3)
- \sum_{n=0}^\infty \int_{p}^{ 2n+5}  \phi^a(t,p_1) \nonumber \\
&\times & \frac{p_2+p_3}{2}
\cdot\frac{p_4+p_5}{2}
\prod_{j=1}^{n+2} \vec\phi(t,p_{2j})\cdot\vec\phi(t,p_{2j+1}) ,
\end{eqnarray}
where we define
\begin{eqnarray}
\int_p^n &\equiv& \prod_{i=1}^n \int \frac{d^2 p_i}{(2\pi)^2} \hat\delta\left(\sum_{i=1}^n p_i - p\right),  \quad
\hat\delta(p) \equiv (2\pi)^2\delta^{(2)}(p) .
\end{eqnarray}
Using the ansatz \eref{eq:sol},  the left hand side is expressed as
\begin{eqnarray}
L^a(t,p) &=& e^{-p^2t} \left[  \dot f(t) \sum_{n=0}^\infty :X_{2n+1}^a(\varphi, p,t) : +f(t) \sum_{n=1}^\infty : \dot X_{2n+1}^a(\varphi, p,t) : \right] ,
\end{eqnarray}
and a similar expression is obtained for the right hand side.


At the leading order of the large $N$ expansion, we obtain
\begin{eqnarray}
\langle L^a(t,p) {\cal O}_1 \rangle &=&  e^{-p^2 t} \dot f(t) \langle \varphi^a(p) {\cal O}_1 \rangle, \\
\langle R^a(t,p) {\cal O}_1 \rangle &=& \lambda e^{-p^2 t} f^3(t) I(t) \langle \varphi^a(p) {\cal O}_1 \rangle + O(1/N), 
\end{eqnarray}
where 
\begin{eqnarray}
I(t) &=& \int \frac{d^2q}{(2\pi)^2}\frac{q^2}{q^2+m^2} e^{-2q^2 t} .
\end{eqnarray}

The gradient flow equation that $ \langle  L^a(t,p) {\cal O}_1 \rangle = \langle R^a(t,p) {\cal O}_1 \rangle $ implies
\begin{eqnarray}
\dot f(t) &=& \lambda f^3(t) I(t), 
\end{eqnarray}
which can be solved as
\begin{eqnarray}
f(t) &=&  \sqrt{\frac{\log(1+\Lambda^2/m^2)}{{\rm Ei}(-2t(\Lambda^2+m^2))-{\rm Ei(-2tm^2)}}} e^{-m^2 t},
\end{eqnarray}
where ${\rm E_i(x)}$ is the exponential integral function defined by ${\rm Ei}(-x) =\int dx \frac{e^{-x}}{x} $.

We can show the finiteness of the flowed field two point function without field renormalization
non-perturbatively at the leading order of the large $N$ expansion, where the two point function is given by
\begin{eqnarray}
\langle \phi^a(t_1, p_1) \phi^b(t_2, p_2) \rangle &=& f(t_1) f(t_2)e^{-p_1^2 t_1}e^{-p_2^2 t_2}\langle \varphi^a(p_1) \varphi^b(p_2) \rangle
\nonumber \\
&=& \frac{f(t_1)f(t_2) \lambda}{N}\delta^{ab}\hat{\delta}(p_1+p_2)\frac{e^{-p_1^2(t_1+t_2)}}{p_1^2+m^2} .
\end{eqnarray}
Since $\lambda f(t_1) f(t_2)$ in the limit $\Lambda\rightarrow\infty$ becomes
\begin{eqnarray}
\lim_{\Lambda\rightarrow\infty} \lambda f(t_1) f(t_2) &=& 4\pi \frac{e^{-m^2(t_1+t_2)}}{ \sqrt{- {\rm Ei} (-2t_1 m^2)}\sqrt{- {\rm Ei} (-2t_2 m^2)}},
\end{eqnarray}
the two point function for the bare field $\phi$ is shown to be finite as
\begin{eqnarray}
\langle \phi^a(t_1, p_1) \phi^b(t_2, p_2) \rangle&=& 
\frac{4\pi e^{-(p_1^2+m^2)(t_1+t_2)}\delta^{ab}\hat{\delta}(p_1+p_2)}{ N\sqrt{- {\rm Ei} (-2t_1 m^2)}\sqrt{- {\rm Ei} (-2t_2 m^2)}} \frac{1}{p_1^2+m^2}
\label{eq:grad-2pt}
\end{eqnarray}
unless $t_1 t_2 = 0$. 
Since any $n$-point functions of the $d=2$, $O(N)$ nonlinear sigma model
can be expressed in terms of the two point function in the large $N$ limit, the flowed $d=2$, $O(N)$ nonlinear sigma model is finite non-perturbatively in this limit .

The two point function of flowed fields was independently analyzed in Ref.\cite{Makino:2014cxa} by the different method, and the result turns out to be consistent with ours.

\section{Discussion}

The gradient flow equation has the attractive property that any composite operators constructed from flowed fields at finite $t$ 
are finite without additional renormalization. Therefore,  it is worth extending the method to various systems.
 In this report , we proposed the generalized gradient flow equation for quantum fields theories with nonlinearly realized symmetries, and apply it to the $\mathcal{N}=1, d=4$, $SU(N)$ super Yang-Mills theory and the $d=2, O(N)$ nonlinear sigma model. 
 
There are various directions in future researches.  
First of all, it is important to investigate the finiteness for  the flowed field in the super Yang-Mills theory, as in the case of Yang-Mills theory\cite{Luscher:2011bx}.
It is also interesting to extend this method to $\mathcal{N}=2$, or $4$ supersymmetric theories.
Using the next to leading order results in the $d=2, O(N)$ nonlinear sigma model\cite{Aoki:2014dxa}, we may explicitly examine the finiteness of the connected four point function for the flowed field.
A different aspect of the gradient flow equation has already been studied in Ref.\cite{Aoki:2015dla}. 
Finally  it may be important to understand a possible relation between the gradient flow equation and the exact renormalization group equation.

\acknowledgments
We would like to thank Hiroshi Suzuki for useful discussions. We also thank Masanori Hanada, Satoshi Yamaguchi,  Koji Hashimoto and Akinori Tanaka for fruitful discussions and comments. 
This work was supported in part by Grant-in-Aid for JSPS Fellows Grant Number 25$\cdot$1336, the Grant-in-Aid of the Japanese Ministry of Education (Nos. 25287046, 26400248), MEXT SPIRE,  JICFuS,  and US DOE grant de-sc0011941.

\end{document}